\documentstyle[aps,preprint,12pt]{revtex}
\begin{document}
\draft
\title {Novel Scaling Relation of the Energy Spacing Distribution
        in Quantum--Hall Systems}
\author{Imre Varga$^{1,2}$, Yoshiyuki Ono$^3$, Tomi Ohtsuki$^4$, and
J\'anos Pipek$^1$}
\address{$^1$Department of Theoretical Physics, Institute of
Physics, Technical University of Budapest, H--1521 Budapest,
Hungary}
\address{$^2$Condensed Matter Research Group of the Hungarian Academy
of Sciences, Technical University of Budapest, H--1521 Budapest,
Budafoki \'ut 8, Hungary}
\address{$^3$Department of Physics, Toho University, Miyama 2--2--1,
Funabashi, Chiba 274, Japan}
\address{$^4$Department of Physics, Sophia University, Kioi--cho 7--1,
Chiyoda--ku, Tokyo 102, Japan}
\maketitle
\begin{abstract}
    The shape analysis of the energy spacing distribution $P(s)$
    obtained from numerical simulation of two dimensional disordered
    electron systems subject to strong magnetic fields is performed.
    In the present work we reanalyze the data obtained in a previous
    publication. Special moments of the $P(s)$ function related to
    R\'enyi--entropy differences show a {\it novel} scale invariant
    relation that is attributed to the presence of one--parameter
    scaling. This relation seems to show both deviations and
    universality depending on which Landau--band is considered and
    whether the disorder is correlated or uncorrelated. Furthermore,
    our analysis shows the existence of an {\it huge, however,
    irrelevant length scale} in the case of the second
    lowest Landau--band and no disorder correlations that completely
    disappears with the introduction of disorder correlations on the
    range of one magnetic length.
\end{abstract}
\newpage
\narrowtext

    The study of the statistical properties of spectra of disordered
    mesoscopic systems attracted much attention during the past
    decade. One of the main issues of this research became the study
    of the disorder induced localization--delocalization transition in
    $d\geq 2$. In the metallic regime while the correlation length is
    much larger than the system size the states are expected to be
    delocalized producing spectral fluctuations that are well
    described by the random matrix theory (RMT). \cite{Meh} The most
    prominent feature of this theory is the presence of level
    repulsion depending on the global symmetry of the system
    (orthogonal, unitary or symplectic). On the other hand in the
    insulating regime the states are exponentially localized therefore
    level clustering occurs. In the critical regime, however, the
    states are multifractal objects still showing similarities to RMT
    behavior although new statistics is expected reflecting the heavy
    spatial fluctuations of the eigenstates. The main purpose of the
    present work is to see how spectral statistics change when
    approaching the transition in quantum--Hall systems.

    One of the tools to analyze the spectral statistics is the study
    of the nearest neighbor energy spacing distribution function
    $P(s)$. This function tends to $P_P(s)=\exp (-s)$ in the
    insulating regime corresponding to Poisson statistics and to
    $P_W(s)=(32/\pi^2)s^2\exp(-4s^2/\pi)$ corresponding to Wigner
    statistics that is a good description of the RMT behavior in the
    metallic regime. In the critical regime the form of $P(s)$ is yet
    unknown.

    The system we study is a simplified version of the model of
    electrons moving in two dimensions subject to a perpendicular,
    strong magnetic field. We refer to all details to Ref. \cite{OOK}.
    Here we only mention that $N\times N$ random Landau--matrices were
    diagonalized and the unfolded spectra $\{x_i \}$ were calculated
    for many samples in order to ensure good statistics. The size of
    the matrix $N$ corresponds to the system size $L=\sqrt {2\pi
    N}\ell$, where $\ell=\sqrt {\hbar/eB}$ is the magnetic length.
    Both the lowest and the second lowest Landau--band are
    considered. In either case the effect of Gaussian disorder
    correlations are studied with a correlation length $\zeta=0$, or
    $\ell$.

    Based on the general understanding of the integer Quantum--Hall
    Effect (QHE) \cite{qhe} the Landau--bands contain localized states
    except in the critical region in the band--center that reduces to
    an extended level at $E_c$ in the thermodynamic limit. In the
    vicinity of $E_c$ one--parameter scaling is expected
\begin{equation}
    Z(E,L)=f(L/\xi),\qquad \xi\sim |E-E_c|^{-\nu}
\label{eq:sc1}
\end{equation}
    where $Z$ is quantity describing the shape of $P(s)$ functions and
    $\xi$ is the localization length, $\nu$ is its exponent. Due to
    the relation between $N$ and $L$ in the representation of the
    unfolded levels the scaling relation reads as
\begin{equation}
    Z(x,N)=\tilde f(x/N^{\delta}),\qquad \nu ={{1}\over{2(1-\delta)}}
\label{eq:sc2}
\end{equation}

    Such quantities that obey the scaling law (\ref{eq:sc2}) have been
    successfully applied in the case of the three dimensional Anderson
    model \cite{VHSP}. They are differences of generalized
    R\'enyi--entropies
\begin{equation}
    -\ln(q)=-\ln\left ({{\mu_1^2}\over{\mu_2}}\right )
    \qquad{\rm and}\qquad
    S_{str}={{\mu_S}\over{\mu_1}}+
            \ln\left ({{\mu_2}\over{\mu_1}}\right)
\label{eq:qs}
\end{equation}
    where $\mu_k=\int s^kP(s)ds$ and $\mu_S=-\int s\ln s P(s)ds$.
    Obviously for the $P(s)$ by definition $\mu_1=1$. These quantities
    have been originally used in the context of wave funtion shape
    analysis (cf. \cite{PV1}) but they are appropriate to study shapes
    of distribution functions in general, e.g. the $P(s)$ \cite{imi}.
    Our parameters may be appropriately rescaled to
\begin{eqnarray}
    \tilde Q={{-\ln(q)+\ln(q_W)}\over {-\ln(q_P)+\ln(q_W)}}
    \qquad \mbox{and}\qquad
    \tilde S={{S_{str}-S_W}\over {S_P-S_W}}
\label{eq:resc}
\end{eqnarray}
    where $_P$ and $_W$ refer to the corresponding limiting cases the
    Poisson-- and the Wigner--statistics. In this way we obtain $0\leq
    \tilde Q,\tilde S\leq 1$, where the value 0 corresponds to
    the Poisson--limit and 1 to the Wigner--limit. In Fig. (\ref{f:1})
    we present the results already in terms of the rescaled variable
    $x/N^{\delta}$. For the lowest Landau--band ($n=0$) the critical
    exponent $\nu=2.4$ is consistent with the literature, on the other
    hand for $n=1$ it is 7.1. The latter value is obviously wrong as
    it has been calculated assuming a simple one--parameter scaling.
    In the figure our parameters show almost full metallic behavior
    ($\tilde S\approx \tilde Q\approx 1$) for $n=1$ and $\zeta=0$.
    This effect should be attributed to the existence of a huge,
    albeit irrelevant length scale. \cite{bh} It is also clear that
    the introduction of disorder correlations makes this pathological
    behavior disappear. The critical indeces obtained from $\delta$ in
    Fig. (\ref{f:1}) are in the range of 1.7--2.4 and show similar
    trends.

    Figure (\ref{f:2}) shows the data on a $\tilde S$ vs. $\tilde Q$
    plot. As a comparision we also plot the evolution of our
    parameters for the interpolating $P_{\alpha}(s)$
\begin{equation}
    P(s)=c(1+\alpha )\,s^{2\alpha}\,\exp (-c\,s^{1+\alpha}),
\label{eq:kl}
\end{equation}
    where $c$ is obtained from $\int sP(s)ds=1$, as $\alpha $ runs
    from 0 (Poisson) to 1 (Wigner) limits.

    It is remarkable that the data for $n=0,1$ and $\zeta =0$ fall on
    more or less the same curve, and so it does for $\zeta =\ell$,
    however, these two cases ($\zeta =0,1$) seem to deviate at least
    in the band--tails. On the other hand as for the band--center,
    the data close to $\tilde Q\to 1$ and $\tilde S\to 1$ fall on
    the curve corresponding to Eq. (\ref{eq:kl}). It seems that
    further analytical as well as numerical work is required in order
    to clarify this kind of correlation.


    Fruitful discussions with Bodo Huckestein is gratefully
    acknowledged. Financial support from Orsz\'agos Tudom\'anyos
    Kutat\'asi Alap (OTKA), Grant Nos. T014413/1994, T021128/1996 and
    F024135/1997 is gratefully acknowledged.

%


\begin{figure}
\caption{\label{f:1}
    The rescaled quantities $\tilde S$ (solid symbols) and $\tilde Q$
    (open symbols) as a function of the rescaled variable
    $x/N^{\delta}$ for different Landau--bands ($n=0,1$) and disorder
    correlations $\zeta =0,\ell$. The data represent different system
    sizes: $N=200$ (squares), 400 (triangles), 600 (diamonds), and 800
    (circles). ($a$) is for $n=0$ and $\zeta =0$, ($b$) is for
    $n=0$ and $\zeta =\ell$, ($c$) is for $n=1$ and $\zeta =0$, ($d$)
    is for $n=1$ and $\zeta =\ell$}
\end{figure}

\begin{figure}
\caption{\label{f:2}
    Correlations between $\tilde S$ and $\tilde Q$. The symbols
    represent different system sizes as in Fig. \protect\ref{f:1}.
    Open (solid) symbol stands for the lowest (second) Landau--band.
    ($a$) no disorder correlations, ($b$) with $\zeta =\ell$ disorder
    correlations. The continuous curve represents the relation for the
    $P(s)$ in Eq. (\protect\ref{eq:kl}).}
\end{figure}

\begin{references}

\bibitem{Meh}M. L. Mehta, {\it Random Matrices} (Academic Press,
    Boston, 1991).

\bibitem{OOK}Y. Ono, T. Ohtsuki, and B. Kramer, J. Phys. Soc. of
    Japan, {\bf 65} (1996) 1734, see references therein.

\bibitem{qhe}M. Janssen, O. Viehweger, U. Fastenrath and J. Hajdu,
    {\it Introduction to the Theory of the Integer Quantum Hall
    Effect} (VCH, Weinheim, 1994); B. Huckestein, Rev. Mod. Phys. {\bf
    67}, 357 (1995).

\bibitem{VHSP}I. Varga, E. Hofstetter, M. Schreiber, and J. Pipek,
    Phys. Rev. B {\bf 52} 7283, (1995); I. Varga, E. Hofstetter, and
    J. Pipek, to be published.

\bibitem{PV1}J. Pipek and I. Varga, Phys. Rev. A {\bf 46} 3148,
    (1992).

\bibitem{imi}I. Varga, Ph. D. Thesis, Technical University of
    Budapest (1993), unpublished; J. Pipek and I. Varga, Intern. J. of
    Quantum Chem., {\bf 64} 85, (1997).

\bibitem{bh}B. Huckestein, Phys. Rev. Lett. {\bf 72} 1080, (1994).
\end{references}
\end{document}